\documentclass[hyper]{JHEP3}

 \usepackage{amssymb}
 \usepackage{amsmath}
 \def\nn{\nonumber}
 \newcommand{\cross}{\times}
 \newcommand{\bchi}{{\mbox{\boldmath $\chi$}}}
 \renewcommand\Im{\hbox{{\rm Im}}\,}
 \renewcommand\Re{\hbox{{\rm Re}}\,}

\newcommand{\ft}[2]{{\textstyle\frac{#1}{#2}}}

\title{Almost BPS black holes}

\author{Kevin  Goldstein  and Stefanos Katmadas
\\ {Institute for Theoretical Physics and Spinoza Institute,\\ Utrecht University,
  Utrecht, The Netherlands}
\\ \email{k.goldstein [at] uu.nl} , \email{s.katmadas  [at] uu.nl}} 

\abstract{
  We study non-BPS black hole solutions to ungauged supergravity with 8 supercharges
  coupled to vector multiplets in four and five dimensions. We identify a large class of
  five dimensional non-BPS solutions, which we call ``almost BPS'', that are
  supersymmetric on local patches and satisfy a first order flow governed by harmonic
  functions. By dimensional reduction, they give rise to new non-BPS solutions in four
  dimensions. These solutions allow for some nontrivial asymptotic moduli and multiple
  centres, similar to their globally supersymmetric cousins. We explicitly discuss a
  single centre and a two centre example.
}
\preprint{{arXiv:0812.4183 [hep-th]} \\{SPIN-08/59} \\ {ITP-UU-08/77} }

\keywords{Black Holes in String Theory, Supergravity Models}

\begin{document}

\section{Introduction and Overview}
\label{intro}

BPS black hole attractors, \cite{9508072,9602111,9602136}, due to a high degree of
supersymmetry, are in some sense, the simplest black holes one can study.  Consequently,
it is not surprising that much progress in understanding the physics of black holes has
come from studying these objects.  The next simplest case one could choose to study is
that of non-BPS black attractors \cite{9702103,0507096}. Although these solutions break
supersymmetry, they seem to generically have similar near-horizon bosonic symmetries to
their BPS cousins \cite{0506177}.  While much can be learnt from  the near horizon
behaviour, an obvious drawback is that, features of the full solution that cannot survive
the near horizon limit are beyond reach. Such features include, the existence and
properties of an interpolating solution to an asymptotic space-time, multi-centred
solutions and associated split attractor flows.  Unfortunately, in the a priori absence of
first order equations, finding the full flow is
challenging\footnote{A frame work for addressing the question of the existence of the full
  flow in the absence of first order equations is laid out in \cite{0507096}.}.

It is known that in four dimensional, $\mathcal{N}>2$ supergravity there is a formal
analytic continuation of charges that can transform certain BPS to non-BPS solutions
\cite{0511117}. This is related to the fact that, at the horizon, the corresponding
solutions to the attractor equations, are related by a sign change. For BPS attractors,
one can obtain the full interpolating solution, from the horizon to the boundary of
space-time, by replacing the charges in the attractor values of the scalars with harmonic
functions \cite{9703101,9704147,9705169,Breitenlohner:1987dg,0012232}. At the level of two
derivative gravity, these solutions satisfy first order flow equations
\cite{9702103,9807087,0005049}. By analytic continuation of the harmonic functions rather
than just the charges, this procedure can be generalised, in certain cases, to non-BPS
attractors \cite{0606263}.  However, these cases are far from being generic --- for
instance once one turns on asymptotic B-fields, which are encoded in certain asymptotic
moduli, the trick of analytically continuing the harmonic functions does not
work --- another approach is needed. It has been found that non-BPS first-order flow equations can also be
found using the so-called ``fake superpotential'' formalism
\cite{0702088,0706.0712, 0706.3373}.  Recently, the addition of nontrivial B-fields was
considered in \cite{0706.3373, 0710.4967, 0807.3503}, using the fake superpotential
formalism and direct calculation, resulting in more complicated solutions that indeed do
not fit in the known analytic continuation scheme.

In this work, we develop a method to obtain non-BPS solutions from the BPS ones that
extends the results mentioned in the previous paragraph. We focus on non-BPS black holes
in four and five dimensional ungauged supergravity.  For concreteness we use theories with
8 supercharges, but the core of our results can be easily applied to the theories with
more supersymmetry. We extend the simple analytic continuation of charges to an associated
conjugation of the five dimensional BPS equations that describe the full flow. Using this
scheme, we find a subclass of non-BPS solutions which, except for global constraints, obey
the five dimensional BPS equations. Consequently, we call these solutions and the
equations they obey ``almost BPS''.  Our scheme reduces to a sign change for the examples
with trivial B-fields, as in the known analytically continued solutions \cite{0606263},
and furthermore includes the recent solutions of \cite{0706.3373, 0710.4967} with
nontrivial B-fields, as well as novel multi-centre solutions.  This generalises previous
results for first order flows of single centred non-BPS solutions
\cite{0702088,0706.0712,0706.3373}, albeit from a different perspective. Furthermore,
hither to mysterious features of these solutions can be understood as a consequence of
their almost BPS nature.

The main technical advantage in five dimensions is the existence of two independent sets
of BPS solutions when the base space is flat. For a more general base, only one of these
sets is BPS according to the known classification \cite{0209114, 0408122}. The main idea
we use to construct the almost BPS solutions is to exploit the simple fact that any
Gibbons-Hawking space can be trivialised into flat space in local patches. This allows us
to replace the known BPS solution by its would-be-BPS partner on every patch. The result
is a new class of solutions that are BPS locally but not globally. In particular, they are
described by first order equations that can be solved in terms of arbitrary harmonic
functions, very similar to the BPS case.

This construction implies that our solutions are constrained to charges that can be
obtained by sign changes from BPS charge vectors, even though the full solutions are very
different. Thus, several known features of more general non-BPS solutions, such as flat
directions \cite{9708025, 0606108, 0606209, 0706.1667, 0709.3488, 0710.4967} remain out of
reach. Nevertheless, our solutions can be used as seeds to generate new ones through known
dualities. It remains an open problem to investigate what subspace of the full non-BPS
spectrum can be obtained in this way. In special cases with enough symmetries, the most
general non-BPS solution can be generated in this way, as in \cite{0710.4967, 0807.3503}
for the case of the STU model.

Restricting attention to solutions that can be conveniently reduced on a circle, we obtain
four dimensional non-BPS solutions characterised by arbitrary harmonic functions that
allow for some nontrivial asymptotic moduli and multiple centres. When taking the near
horizon limit, our solutions reproduce the known four dimensional non-BPS attractors. From
the five dimensional point of view, the near horizon geometry is a time-like fibration over flat
space, which locally satisfies the BPS equations presented in the next section. This implies
that the near horizon geometries of our solutions are solutions to the attractor
equations. It also allows us to construct stable multi-attractors with the same total
charge as single centre non-BPS attractors whose stability is unclear
\cite{0606263,0511117}. The same situation is found at the asymptotically flat region,
leading to severe constraints on the form of the mass formula.

The crucial property of being locally supersymmetric implies strong constraints on the
behaviour of our solutions in the interpolating region as well. The most important is the
existence of a first order flow for the scalars governed by the almost BPS equations. As
far as we know, by U-duality, this includes all known cases of four dimensional non-BPS
first order flows in $\mathcal{N}=2,4,6,8$ theories and extends it to the multi-centre
case.

This paper is organised as follows. In section \ref{sec:bg} we give some background
material on five dimensional supergravity, the classification of its BPS solutions and its
reduction to four dimensional $\mathcal{N}=2$ supergravity. Then, in section
\ref{sec:nonBPS} we discuss the doubling of BPS solutions for the case of flat base and
derive the almost BPS equations for a non-flat base. Section \ref{sec:eg}
contains two particular examples that are interesting from a four dimensional point of
view, a single centre and a two centre solution. We conclude in section \ref{sec:conc}
with some extra comments and directions for future study.

\section{5D supergravity and dimensional reduction}
\label{sec:bg}

In this section we include some background material and establish notation. First, we
briefly discuss five dimensional ungauged supergravity coupled to vector multiplets and
the classification of the supersymmetric solutions of this theory. Next, we consider the
dimensional reduction to four dimensions that yields the $\mathcal{N}=2$ supergravity we
are interested in.

\subsection{BPS solutions of five dimensional supergravity}
\label{sec:5d}
We consider five dimensional ungauged supergravity coupled to $n_v$ vector
multiplets. Unless stated otherwise, we adopt the conventions of \cite{0408122}.  Using a
positive signature metric, the bosonic action takes the form \cite{Gunaydin:1983bi}:
\begin{multline}\label{5dact}
  S = \frac {1}{ 2 \kappa_{5}} \int\!\sqrt{-g}\,d^5x \bigg( R  -\ft{1}{2} Q_{IJ} F_{\mu \nu}^I 
    F^{J \mu \nu} -Q_{IJ} \partial_\mu X^I  \partial^\mu X^J
    \\-\ft {1}{24} C_{IJK} F^I_{ \mu \nu} F^J_{\rho\sigma} A^K_{\lambda} \bar\epsilon^{\mu\nu\rho\sigma\lambda}\bigg),
\end{multline} 
where $\bar\epsilon^{\mu\nu\rho\sigma\lambda}$ is the completely antisymmetric Levi-Civita
{\em tensor} with $|\bar\epsilon^{\mu\nu\rho\sigma\lambda}|=1/\sqrt{-g}$ (the orientation
will be defined below). The indices, $I, J, K\ldots = 1,\ldots, n_v$, label the vector
multiplets and $C_{IJK}$ is a totally symmetric constant tensor. The scalars $X^I$ satisfy
the constraint:
\begin{equation}
  \label{conda}
  X_I X^I =1\,,\qquad X_I \equiv \frac 16 C_{IJK} X^J X^K \ ,
\end{equation}
so that only $n_v-1$ of them are independent. We denote the independent scalars as,
$\chi^{a}$, where $a=1,\ldots,n_{v}-1$. The scalar dependent coupling matrix, $Q_{IJ}$, has the form:
\begin{equation}\label{scamatr}
  Q_{IJ} = \frac 9 2 X_I X_J -\frac 1 2 C_{IJK} X^K \ .
\end{equation}
For simplicity, we will restrict our attention to the case where the scalars take values
in a symmetric space. One can then use the identities:
\begin{eqnarray}
  \label{eqn:symm} 
  \frac 4 3 \delta_{I (L} C_{MPQ)} &=&
  C_{IJK} C_{J' (LM} C_{PQ) K'} \delta^{J J'} \delta^{K K'} \,,\\ 
  X^I&=&\frac{9}{2} C^{IJK} X_JX_K\,, \label{eqn:symm2}
\end{eqnarray}
where $C^{IJK}\equiv \delta^{II'}\delta^{JJ'}\delta^{KK'}C_{I'J'K'}$ and the constraint in
\eqref{conda} was used in obtaining (\ref{eqn:symm2}).


The supersymmetric solutions of this theory have been classified in \cite{0209114,0408122}
and fall in two distinct classes, depending on whether the vector,
$\bar{\epsilon} \gamma^\mu \epsilon$, constructed from the Killing spinor, $\epsilon$, is
time-like or null. We will restrict our attention to the time-like case, in which the BPS
conditions imply that the metric is a time-like fibration over a hyper-K\"ahler base
space, which we briefly summarise below.  

Given a hyper-K\"ahler 4-manifold with a metric, $h_{mn}$, we choose the associated
complex structures, $X^{(i)}$, as anti-self-dual and assume for the moment that they are
unique. Given these data, the metric and gauge fields of a supersymmetric solution can be
written locally as:
\begin{eqnarray}\label{bps_metric}
  ds^2&=& -f^2(dt+\omega)^2 + f^{-1}h_{mn}dx^m dx^n \nn\\
  {}F^I &=&  d(X^I e^0) + \Lambda^{I} .
\end{eqnarray}
Here, $e^0=f(dt+\omega)$, the indices, $m,n,\ldots=1,\ldots,4$, label coordinates on the
base, the $\Lambda^I(x^{m})$ are arbitrary closed self-dual forms on the base, and $f>0$,
is assumed to be a globally defined function. A positive orientation is chosen using
$e^0\wedge \eta$ as the volume form, where $\eta$ is a positive orientation on the base
manifold.

Once the $\Lambda^I$'s are chosen, the function $f$ and the $X^I$ are determined by
solving a Poisson equation on the base:
\begin{eqnarray}\label{bps_scal}
  \Delta \left( f^{-1} X_I \right)=\frac1{12} C_{IJK}\Lambda^{J}_{mn} \Lambda^{ mn\,K}\,,
\end{eqnarray}
where $\Delta$ is the Laplacian on the hyper-K\"ahler manifold. Finally, the one-form,
$\omega$, is determined by solving:
\begin{eqnarray}\label{bps_omeg}
  f d\omega = G = G^+ + G^-, \qquad X_I \Lambda^{I}=- \frac23 G^+ \,,
\end{eqnarray}
where $G^{\pm}$ are self- and anti-self-dual forms on the hyper-K\"ahler base. These BPS
solutions are invariant under an $SU(2)$ subgroup of the base space isometry group and, in
an orthonormal basis, their respective Killing spinors satisfy
$\gamma^0\epsilon=\mathrm{i}\epsilon$. Note that if our assumption of the uniqueness of the complex
structures is not true, there is extra freedom on the BPS solutions that can be written
down, changing the above equations. This will be seen explicitly in the following, but for
the moment we continue to assume uniqueness.  

In this work we are interested in solutions with a four dimensional interpretation, so we
will demand that the base space has a compact isometry along which we can perform
dimensional reduction. Under the assumption that the associated Killing vector is
triholomorphic (i.e. leaves the complex structures invariant), and is a symmetry of the full
solution, the base space can only be a Gibbons-Hawking space \cite{Gibbons:1987sp}. In
this case the above equations simplify enough to be solved explicitly \cite{0408122}.  For
a Gibbons-Hawking space \cite{Gibbons:1979zt}, which is itself a fibration over a flat
Euclidean base, the metric, $h_{mn}$, can be written:
\begin{eqnarray}\label{GH_metric}
  h_{mn}dx^m dx^n &=& H^{-1} \left( d\psi + \chi_i dx^i \right)^2 + H \delta_{ij}dx^i dx^j, \\
  \nabla \cross \bchi &=& \nabla H. \label{GH_metric2}
\end{eqnarray} 
Here, $\nabla$ is the standard vector derivative on the Euclidean 3-space,
$\mathbb{R}^{3}$, with coordinates $x^i\,, i,j=1,2,3$, $H(x^{i})$ is a harmonic function
on $\mathbb{R}^{3}$ and $ 0 \leq\psi\leq 4 \pi$. The isometry group is $SU(2)\times U(1)$,
where the $U(1)$ is generated by the Killing vector $\partial/\partial \psi$. We will
consider the specific examples of flat space ($H=1$ or $H=1/|{\bf x}|$) and Taub-NUT space
($H=h^0 + p^0/|{\bf x}|$).

The complex structures associated with \eqref{GH_metric} are given by
\begin{equation}\label{cpmstr}
  X^{(i)} = \left( d\psi + \chi_j dx^j \right) \wedge dx^i - \frac{1}{2}
  H \epsilon_{ijk} dx^j \wedge dx^k.
\end{equation}
Using (\ref{GH_metric2}), one can easily see $dX^{(i)}=0$. Imposing their
anti-self-duality fixes the orientation of the base space so that the volume form is
\begin{equation}
  H d\psi \wedge dx^1 \wedge dx^2 \wedge dx^3.
\end{equation}
The complex structures \eqref{cpmstr} are globally defined and, in all but one case,
unique. The exception is flat space (in coordinates such that $H=1$, $\chi=0$), in which
case one can also choose the opposite relative sign in \eqref{cpmstr} and the forms,
$X^{(i)}$, remain closed. It follows that for flat space there are two triplets of complex
structures: one self-dual and one anti-self-dual. This observation implies an enlargement
of the set of BPS solutions in that case, to be discussed in section~\ref{sec:nonBPS}.

The explicit BPS solution for a Gibbons-Hawking base can be described in terms of $H$
and an additional $2 n_v + 1$ harmonic functions \cite{0408122}. The self-dual forms
$\Lambda^I$, defined in (\ref{bps_metric}), can be written as:
\begin{equation}
  \Lambda^I = -\frac1 2 (d\psi + \chi) \wedge
  (W^I{}_j) dx^j -\frac1 4 H \epsilon_{ijk}
  (W^I{}_k) dx^i \wedge dx^j\,,\qquad W^I{}_i=\partial_i\left(\frac{K^I}{H}\right)\,,
\end{equation}
where the $K^I$ are arbitrary harmonic functions related to the magnetic charges of the
solution. Given these functions, the rest of the solution is found by solving
\begin{eqnarray}
  \label{eqn:fscal}
  f^{-1} X_I &=& \frac1 {24} H^{-1} C_{IJK} K^J K^K +L_I\,, \\
  \omega_5 &=&- \frac1{48}H^{-2}C_{IJK} K^I K^J K^K- \frac3 4 H^{-1} L_I K^I +M\\
  \nabla \times {\hat{\omega}} &=& H \nabla M - M \nabla H +\frac3 4
  (L_I  \nabla K^I - K^I \nabla L_I) \label{omeg_BPS}\,.
\end{eqnarray}
Here, the one-form $\omega$ is decomposed as
\begin{eqnarray}\label{omega}
\omega_m dx^m= \hat\omega_i dx^i+ \omega_5(d\psi + \chi)\,,
\end{eqnarray} 
and $L_I$, $M$ are arbitrary
harmonic functions associated with the electric charges and the angular momentum along the
$\psi$ direction respectively.

\subsection{4D/5D connection}
In the following, we will be interested in the properties of the solutions to the four
dimensional theory obtained by reducing the theory \eqref{5dact} on a circle. In order to
make the connection clear, we present some of the relevant formulae for this reduction
following \cite{0506251,0701176}, but using slightly different conventions.

The relevant ansatz for the reduction is written as:
\begin{eqnarray}\label{red_ans}
  ds^2 &=& e^{2 \phi} \, ds^2_{(4)} + e^{- 4 \phi} \,(d \psi - A^0_{(4)})^2 \;,
  \nonumber\\
  A^I &=& A_{(4)}^I + C^I \, (d \psi - A_{(4)}^0) \;,\nonumber\\
  {\hat X}^I &=& e^{- 2 \phi} \, X^I\;,
\end{eqnarray}
where $ds^2_{(4)}$ is the four dimensional line element and the $C^I$, ${\hat X}^I$ will
make up the four dimensional complex scalars. Here, the coordinate along the circle,
$\psi$, runs over $ 0 \leq\psi\leq 4 \pi$.  This fixes the four dimensional Newton
constant: $G_4 = G_{5}/4 \pi$.

Reducing the action in \eqref{5dact} along a circle \cite{0701176}, one obtains a four
dimensional ${\mathcal N}=2$ supergravity characterised by the pre-potential:
\begin{eqnarray}
  F(Y) = - \frac1{12} \, \frac{C_{IJK} \, Y^I Y^J Y^K}{Y^0} \;,
  \label{prepotential}
\end{eqnarray}
where the $Y^A$ are $n_{v}+1$ complex scalar fields ($A=0, I$). These scalars are also
subject to a constraint that makes one of them redundant. We solve it by working
throughout in special coordinates:
\begin{equation}\label{4d_scal}
  z^I = \frac{Y^I}{Y^0}= C^I+ \mathrm{i} {\hat X}^I \;.
\end{equation}
The explicit four dimensional bosonic action that results from the reduction is
\cite{deWit:1984pk, deWit:1984px}:
\begin{multline}\label{4dact}
  S_4 = \frac{1}{2 \kappa_4} \int d^4 x  \sqrt{-g}  \bigg( R - 
    2 g_{I \bar J}  \partial_{\mu} z^I \partial^{\mu} 
    {\bar z}^J + \ft{1}{2}  {\Im} {\cal N}_{AB}  F_{\mu \nu}^A 
    F^{B \mu \nu} 
    \\ - \ft{1}{2} 
    {\Re} {\cal N}_{AB}  F_{\mu \nu}^A *{F}^{B \mu \nu} 
  \bigg),
\end{multline}
where $*{F}^{A\, \mu\nu} = \frac12 \epsilon^{\mu\nu\rho\sigma} F^{A}_{\rho\sigma}$ with
$\epsilon^{0123} = 1/\sqrt{-g}$. The four-dimensional gauge couplings, ${\cal N}_{AB}$,
are given by
\begin{eqnarray}\label{gau_coupl}
  \mathcal{N}_{AB}
  =\bar F_{AB}+ 2\, \mathrm{i}\, \frac{\Im F_{AC} \, \Im F_{BD} \, Y^C Y^D}{\Im F_{FE} 
    \, Y^F Y^E}\;,
\end{eqnarray}
where $F_A = \partial F/\partial Y^A \,,\, F_{AB} = \partial^2 F/\partial Y^A
\partial Y^B$. Finally, the moduli space metric, $g_{I \bar J}$, is:
\begin{equation}
  g_{I \bar J}
  = \frac{\partial}{\partial z^I} \frac{\partial}{\partial {\bar z}^J} K 
  \,, \qquad {\rm e}^{-K (z, {\bar z})} = \frac{\mathrm{i} }{12} C_{IJK} (z^I - {\bar z}^I)
  (z^J - {\bar z}^J) (z^K - {\bar z}^K) \;.
\end{equation}

Note that there is an extra gauge multiplet in four dimensions, represented by the zeroth
index, which comes from the reduction of the five dimensional graviton supermultiplet. The
scalars are complexified in the reduction, with the $n_v$ pseudoscalars $C^I$ coming from
the gauge fields paired with the $n_v$ physical scalars $X^I$ and the Kaluza-Klein scalar
$\phi$ as in \eqref{4d_scal}. In string theory language, the $C^I$ are interpreted as
components of the B-field reduced to four dimensions.

In contrast to the above, finding the explicit relation between the charges of a solution
to the five dimensional theory and its four dimensional reduction can be subtle,
especially in the presence of magnetic charges. Since, in the following, we are interested in the four
dimensional interpretation,  we will be dealing only with four dimensional
electric and magnetic charges, using the natural formulae:
\begin{eqnarray}
  q_A\equiv(q_0\,,q_I)&=&\frac1{4\pi}\int_{S^2}\left( \Im{\cal N}_{AB}*\! F^B- \Re{\cal N}_{AB} F^B\right)\,,\label{el_chsr}\\
  p^A&\equiv&(p^0\,,p^I)=\frac1{4\pi}\int_{S^2} F^A\,,\label{mag_char}
\end{eqnarray}
where the sphere encloses the region of interest.

This completes the dictionary between the solutions of \eqref{5dact} and \eqref{4dact} in
the general case. For our purposes though, we will be interested only in five dimensional
solutions that can be written as a time-like fibration over a Gibbons-Hawking base space
as in \eqref{bps_metric} and \eqref{GH_metric}. Under this restriction, there are
significant simplifications. In particular, the Kaluza-Klein scalar, ${\phi}$, and the
four dimensional metric, are written as:
\begin{eqnarray}\label{4dmetric}
  ds^2_{(4)} = - {\rm e}^{2U} (dt + \hat{ \omega}_i \, d x^i)^2
  + {\rm e}^{-2U} d\vec x^2\qquad e^{-4U}=\frac{H^2}{f^2}e^{-4\phi}=H\,f^{-3} -(H\,\omega_5)^2\,.
\end{eqnarray}
Moreover, the Kaluza-Klein gauge potential seen in four dimensions is given by the
expression:
\begin{eqnarray}\label{4dKK}
  A^0_{(4)} =  \omega_5 \, H^2 \, {\rm e}^{4U}\; (dt + \hat{ \omega}_i \, d x^i ) - \chi_i \, d x^i \,,
\end{eqnarray}
where the notation in \eqref{omega} is used for $\omega$. In the following, we will use
the above formulae both explicitly and implicitly, to give a four dimensional
interpretation to our results. More explicit details can be found
in\footnote{ Note that our tensor $C_{IJK}$ is related by a factor of two to the one in
  \cite{0701176}} \cite{0701176}.

\section{BPS and almost BPS solutions}\label{sec:nonBPS}
\subsection{BPS solutions with flat base space}\label{sec:flat_base}
As noted in section \ref{sec:5d}, for the special case of a flat base space there is some
extra freedom in choosing the hyper-K\"ahler complex structures: namely they can be chosen
to be either self- or anti-self-dual. This is a result of the larger group of rotations
for $\mathbb{R}^4$, namely $SO(4)\equiv SU(2)_L\times SU(2)_R$. The two triplets of
complex structures rotate under different linear combinations of $SU(2)_R$ and $SU(2)_L$.

In this special context, there is a second supersymmetric solution one can write down, in
which all the fields are left invariant under the self-dual complex structures
\cite{0209114}. Explicitly, the relevant BPS conditions
\eqref{bps_metric}-\eqref{bps_omeg} for the flat case are modified to:
\begin{eqnarray}\label{bpsmetric2}
  ds^2&=&-f^2(dt+\omega)^2+ f^{-1}h_{mn}dx^m dx^n \,,\\
  {}F^I &=& \pm \Big( d(X^I e^0) + \Lambda^{I} \Big) \,,\\
  X_I \Lambda^{I}&=&-\frac23 G^\pm\,, \label{bpsl2}\\
  \Delta \left( f^{-1} X_I \right)&=&\frac1{12} C_{IJK}\Lambda^{J}_{mn} \Lambda^{ mn\,K}\,. \label{lapl}
\end{eqnarray}
Here, $\Lambda^I$ are self-dual forms on the base for the upper sign and anti-self-dual
for the lower sign. These pairs of BPS solutions are aligned with the two complex
structures in the sense that they are invariant under the corresponding $SU(2)$ subgroups
of the $SO(4)$ isometry group and, in an orthonormal basis, their respective Killing
spinors satisfy $\gamma^0\epsilon=\pm\mathrm{i}\epsilon$. Some examples of supersymmetric pairs were
written down in \cite{0209114}, using right- and left- invariant one-forms on $S^3$.

An interesting property of these BPS pairs is that when reducing to four dimensions along
a circle, it is not possible to retain both of them in the BPS spectrum. This is because
any reduction ansatz can only respect one $SU(2)\equiv SO(3)$ isometry, while the two sets of
BPS solutions respect a different $SU(2)$ in five dimensions. Thus, choosing a particular
$SU(2)$ to keep in the reduction, the other one is broken and all the BPS solutions
aligned with it are lost as well.  Equivalently, the Killing spinors in one set of
solutions are invariant under the $SU(2)$ that is being broken, while those of the other
set are not. This means that under the reduction, the Killing spinors in one set will be
trivially reduced, but the ones in the other set will be charged under the Kaluza-Klein
$U(1)$, violating the natural assumption of invariance and leading to a non-BPS
solution. A very similar situation was encountered in \cite{Nilsson:1984bj,9704186}, where
it was shown that if one considers the full Kaluza-Klein tower, supersymmetry is
recovered.

This ``supersymmetry without supersymmetry'' effect \cite{ 9704186}, was observed in
\cite{0701176, 0707.0964}, in the context of lifting to five dimensions the near horizon
geometry of a four dimensional non-BPS black hole, which can be written as a solution with
flat base space. In the particular electric case discussed there, one has $\Lambda^I=0$,
$\omega=0$ and $H=p^0/r$ for flat base space with a conical singularity (the coordinate
$\psi$ has a fixed range as in \eqref{GH_metric}). Then the two five dimensional BPS
solutions above are identical with respect to the metric and the scalars, but have
opposite electric charges. The explicit solutions are described by (in our conventions):
\begin{eqnarray}\label{bps_flat}
  ds^2&=&-f^2dt^2+ f^{-1}\left( \frac {p^0}r d x^i d x^i + \frac r{p^0}(d\psi + p^0 \cos{\theta}\phi)^2 \right) \nn\\
  {}F^I &=& \pm d(f X^I)\wedge dt, \qquad  f^{-1} X_I =L_I=\frac23 \frac{q_I}{\,r}.
\end{eqnarray}
Upon reduction to four dimensions preserving the $SO(3)$ symmetry of the directions $x^i$,
they give the BPS and the non-BPS attractor discussed in \cite{0701176, 0707.0964} for the
plus and minus sign
respectively\footnote{If the other $SO(3)$ symmetry of ${ \mathbb R}^4$ is chosen in the ansatz, one
  ends up with the same solutions in four dimensions, but their origins in five dimensions
  are interchanged.}.
From the four dimensional point of view the two solutions are related by analytic
continuation of the charges \cite{0511117}.

It should be noted that this observation does not affect the 4D/5D connection for BPS
solutions as described in \cite{0503217, 0504126}. There, a Taub-NUT base that
interpolates between a five dimensional and a four dimensional solution was used to argue
that the BPS index is the same in four and five dimensions. As Taub-NUT has a unique
triplet of complex structures, there is only one BPS solution for each choice of harmonic
functions. Thus, all asymptotically Taub-NUT BPS solutions in five dimensions are mapped
to asymptotically flat BPS solutions with appropriate charges, under dimensional reduction
\cite{0506251}. The same holds for any Gibbons-Hawking space except flat space.

This is explicitly seen from the asymptotically Taub-NUT extensions of the two attractor
solutions. The solution with the plus sign in \eqref{bps_flat} is compatible with the
anti-self-dual complex structures of $\mathbb{R}^4$ and its asymptotically Taub-NUT
extension is the BMPV black hole \cite{9602065} in the centre of Taub-NUT. In view of the
anti-self-duality of the complex structures of this base, it is a BPS solution
\cite{0209114}. The solution with the minus sign is instead compatible with the self-dual
complex structures of $\mathbb{R}^4$.  Its Taub-NUT extension was constructed in
\cite{0706.3373}, and is non-BPS. In the following we rederive this solution and clarify
the origin of its special properties, as part of a general technique of constructing
non-BPS solutions.

\subsection{Almost BPS solutions} \label{noBPS} In the less restrictive case of a general
Gibbons-Hawking space, one might wonder about the fate of the two different supersymmetric
solutions. This becomes especially interesting in view of the fact that both kinds of BPS
solutions with a flat base can be the near horizon region of (not necessarily BPS) black
holes embedded in a more general space, as mentioned above. The existence of full interpolating
Taub-NUT solutions for the two attractors above, only one of which is BPS, suggests that the two different 
solutions might survive as BPS/non-BPS pairs in this case.

Indeed, it is straightforward to show that even if the base is not flat, both expressions in
\eqref{bpsmetric2}-\eqref{bpsl2} solve the equations of motion. An outline of this
calculation can be found in Appendix \ref{feom}. For a general base, the anti-self-duality of the hyper-K\"ahler structures 
allows only for the one with the upper sign to be supersymmetric, whereas the other one is not. 

An intuitive picture of the relation between the two solutions can be given for a Gibbons-Hawking 
base space.  The base space is a $U(1)$ bundle over $\mathbb{R}^3$, so it can be trivialised into $\mathbb{R}^4$ 
by a suitable choice of coordinates on any local patch. One then has a choice between self-dual or 
anti-self-dual complex structures on every such patch as before, so that both expressions in 
\eqref{bpsmetric2}-\eqref{bpsl2} constitute BPS solutions.

By extending to the full base space, only the anti-self-dual structures 
on local patches can be integrated to the unique global complex structures \eqref{cpmstr}.
In contrast, the local self-dual structures can be integrated to the almost hyper-K\"ahler structures:
\begin{equation}\label{alm_cpmstr}
  \tilde X^{(i)} = \left( d\psi + \chi_j dx^j \right) \wedge dx^i + \frac{1}{2}
  H \epsilon_{ijk} dx^j \wedge dx^k,
\end{equation}
that are globally defined, but not integrable: $d\,\tilde X^{(i)}\!\neq 0$. The existence of the
forms \eqref{alm_cpmstr} allows one to construct globally defined fields by aligning local solutions on every patch with 
the appropriate restriction of these structures\footnote{Note that the restriction of these forms on a patch 
is transformed to a constant by the coordinate transformation that trivialises the patch. It is 
the noncompatibility of these local coordinate transformations that makes the global forms not 
integrable. This is also what prohibits the existence of a corresponding global Killing spinor.}.
It is then clear why both signs in \eqref{bpsmetric2} provide a solution to the 
equations of motion, since they can be viewed as constructed locally from BPS solutions aligned with the forms in \eqref{alm_cpmstr}. 
The difference is that the one with the upper signs is compatible with the global complex 
structures and is a global BPS solution. The second solution fails to be supersymmetric only
due to a global obstruction, providing an example of a nonsupersymmetric solution with the
peculiar property of admitting four supercharges on local patches. In fact, it is expected
to have all the local properties of a BPS solution, which are behind most of the
computational simplifications in that case. This property, which is based on the existence
of an almost hyper-K\"ahler structure, motivates the nickname, ``almost BPS'', for these
solutions.  By the same argument, our non-BPS solutions are supersymmetric on local 
patches for a hyper-K\"ahler base more general than Gibbons-Hawking if there exists a globally 
defined almost hyper-K\"ahler structure.

Here, we will restrict to Gibbons-Hawking base spaces for simplicity. In this case, the
almost BPS solutions can be specified through arbitrary harmonic functions as for the BPS
case, following the algorithm in \cite{0408122}. First write
\begin{equation}
  \omega = \omega_5 \left(d\psi + \chi_i dx^i \right) + \hat\omega_i dx^i,
\end{equation}
\begin{equation}\label{2form}
  \Lambda^{I} = \frac{1}{2} A^{I}_i \left(d\psi + \chi_j dx^j \right) \wedge dx^i - \frac{1}{4} H
  \epsilon_{ijk} A^{I}_k dx^i \wedge dx^j,
\end{equation}
for the one-form $\omega$ and the anti-self-dual form in \eqref{bpsmetric2}, where $A^I$
are arbitrary one-forms to be determined.

Using \eqref{GH_metric}, closure of $\Lambda^{I}$ reduces to the relations:
\begin{equation}\label{curl}
  \nabla \cross {\bf A}^I = 0\,, \qquad \nabla \cdot  {\bf A}^I = 0 \,,
\end{equation}
which in turn imply that locally:
\begin{equation}\label{ks}
  {\bf A}^I = \nabla K^I\,,
\end{equation} 
for some harmonic functions $K^I$. The equation for $f^{-1}X_I$ in \eqref{bpsl2} reduces
to
\begin{equation} \label{feq} \nabla^2 \left(f^{-1}X_I\right) = \frac{1}{12} H C_{IJK}
  \nabla K^J \nabla K^K = \frac{1}{24} H \nabla^2\left( C_{IJK} K^J K^K\right)\,,
\end{equation}
which can be solved up to a set of arbitrary harmonic functions $L_I$, given the
$K^I$. Here, we will restrict to solutions of the slightly stronger relation:
\begin{equation} \label{attr} \nabla \left(f^{-1}X_I\right) = \frac{1}{24} \Big( H
  \nabla\left( C_{IJK} K^J K^K \right) - C_{IJK} K^J K^K \nabla H\Big) + \nabla L_I\,,
\end{equation}
even though there might be physically interesting solutions not captured by it. The
advantage of this simplification is that the scalars are governed by a first order flow
very similar to the BPS one.

Finally, we find the conditions on $\omega_5$ and $\hat\omega_i$. Writing out the first
equation in \eqref{bpsl2} using \eqref{bps_omeg} and \eqref{ks} one gets:
\begin{eqnarray} \label{omeg} \nabla \times {\hat\omega} +\nabla\left( H \omega_5\right) =
  \frac32 f^{-1}HX_I \nabla K^I
\end{eqnarray}
Taking the divergence of this gives the integrability condition
\begin{equation}\label{omeg_nab}
  \nabla^2\left( H \omega_5 \right) = \frac32
  \nabla \left(f^{-1}HX_I \right)  \cdot \nabla K^I ,
\end{equation}
which can be solved up to an arbitrary harmonic function $M$, given the solution of
\eqref{attr}. Substituting in \eqref{omeg}, the one-form $\hat\omega$ can be determined up
to a total derivative (removable by a change of coordinates).

Observe that, just like its BPS partner, an almost BPS solution is determined by $2n_v+2$
harmonic functions $H$, $K^I$, $L_I$, $M$, which encode the charges as in section
\ref{sec:5d}. Note however, that if $H$ is such that the base is flat, the BPS/almost BPS
pairs degenerate into the supersymmetric pairs in the previous section. Hence, there are
no asymptotically flat five dimensional nonsupersymmetric solutions in the class described
here. On the other hand there are solutions that asymptote to $\mathbb{R}^3\times S^1$,
allowing for an interpretation as asymptotically flat solutions in four dimensions. That
is the case of interest in this work.

In any case, the almost BPS solutions are a five dimensional analogue of the
four dimensional non-BPS black holes that can be obtained from BPS solutions through a
change of relative signs in charges (and harmonic functions). Comparing
\eqref{feq}-\eqref{omeg_nab} with \eqref{eqn:fscal}-\eqref{omeg_BPS}, shows that they are
related in that way if $K^I=0$, but the general case is much more complicated than a
change of relative signs, as it involves different powers of the harmonic function
$H$. For the special case of flat base, where one can choose $H=1$, the supersymmetric pairs 
are always related by such a sign change, but not if one takes $H=1/r$. The latter choice 
typically leads to genuinely different BPS solutions that are not asymptotically flat, 
as in \cite{0209114}.

Interestingly, by reducing along the $\partial/\partial\psi$ direction, our solutions
give rise to nonsupersymmetric solutions of four dimensional $\mathcal{N}=2$ supergravity
that are more general in several ways than the ones produced by the four dimensional sign
change. In particular, the five dimensional almost BPS solutions allow for some nontrivial
moduli at infinity as in \cite{0706.3373, 0710.4967} and multiple centres, unlike the
simple flip of signs in four dimensions that only works for single centre solutions
without B-fields.

Despite this, our solutions are far from being the most general non-BPS black holes, as
they represent very special points in the duality orbits. For example, we cannot construct
a $D0\!-\!D6$ non-BPS black hole (see e.g \cite{0701176, 0710.4967, 0807.3503}) using the
almost BPS equations, as they were derived from the BPS conditions, that do not allow for
such an object. Moreover, it is known that non-BPS black holes generically exhibit flat
directions all along their flow in both four and five dimensions
\cite{9708025, 0606108, 0606209, 0706.1667, 0709.3488}. This feature is also not captured
by our almost BPS solutions, since the BPS solutions have no flat directions
either. Nevertheless, our solutions can be used as seed solutions to generate new ones
through four dimensional dualities. If there are enough symmetries in a specific theory,
the most general non-BPS solution can be generated in this way, as in
\cite{0710.4967, 0807.3503} for the case of the STU model.

Unfortunately though, the explicit equations are significantly more complicated than in
the BPS case and can not be solved formally as in \cite{0408122} in the most general
case. Nevertheless, the full interpolating solution for any specific choice of harmonic
functions should be readily obtainable by a numerical approach, since the solution is
determined through Poisson equations which are well studied. The rest of this paper is
devoted to analysing examples of almost BPS solutions and their properties.

\section{Examples of almost BPS solutions}
\label{sec:eg}

We now turn to some examples of almost BPS solutions in Taub-NUT. We discuss a single
centred and a two centred example, both of which are very interesting from a four
dimensional perspective. In doing this, we draw on intuition from the study of the known BPS
partners of these solutions. The two cases are largely independent, but share some
common features that are generic for our solutions and follow from their construction.

The most important feature is the special structure of their near horizon and asymptotic
regions.  Indeed, since the near horizon region of our solutions can be written as a
timelike fibration over flat space, it preserves four supercharges by construction.  In
fact, it preserves all eight supercharges, since the near horizon region is maximally
supersymmetric for BPS solutions.  The same holds trivially for the asymptotic region in
our Taub-NUT examples up to compactness of one coordinate, as they asymptote to
$\mathbb{R}^3\times S^1$. This property simplifies the near horizon analysis, but more
importantly implies that the mass of these solutions takes the same form as for its BPS
partner, namely that it is a simple sum of the charges. This suggests that the
constituents are marginally bound. This feature was observed for nonsupersymmetric
solutions of the $STU$ model in four dimensions \cite{0710.4967}, some of which can be
lifted to almost BPS solutions in five dimensions, as we will show below. Note that the
BPS nature of the attractors and the asymptotic region of our solutions is invisible from
a four dimensional perspective, as explained in section \ref{sec:flat_base}.

A second, more intricate, property that also follows from the almost BPS conditions is the
existence of a first order flow for the scalars as in
\cite{0702088,0706.0712,0706.3373}. In fact the function governing this flow can be
obtained from the BPS central charge, that plays this role in the BPS case, using our
recipe. The above statements apply both to the five dimensional and the four dimensional
theory, by dimensional reduction. In the single centred case we will show such an explicit
example, reinterpreting the known solution of \cite{0706.3373} in our framework.

Finally, a technical point is that our equations allow for at least one extra asymptotic
modulus than what can be obtained through the four dimensional sign flip. It is described
by the constant part of the arbitrary harmonic function $M$ in the solution of
\eqref{omeg_nab}. This constant is always left undetermined because only derivatives of
$H\omega_5$ and thus $M$ appear in all expressions. This is a crucial difference from the
BPS equation \eqref{omeg_BPS}, through which the constant part of $M$ is related to other
quantities when cancellation of Dirac-Misner strings is imposed. From a five dimensional
point of view, this constant is interpreted as rotation of the asymptotic region, as in
\cite{0504125, 0706.3373}. From the point of view of the four dimensional theory it is seen as an
asymptotic $B$-field, which allows us to make contact with the purely four dimensional
derivation of \cite{0710.4967} in the single centre case.

\subsection{Non-BPS electric black hole}\label{sing_exa}
In this subsection, we make contact with the single centre solutions in
\cite{0706.3373,0710.4967}. Since these solutions were thoroughly examined in
\cite{0706.3373}, we will be brief, emphasising general aspects. Similar to the BMPV
solution on the BPS side, we choose the harmonic functions as:
\begin{equation}\label{har_fun_one}
  H=h^0+\frac{p^0}{r}\,,\quad K^I=0\,,\quad L_I=l_I+ \frac23\frac{\,q_I}{r}\,,\quad M=-b\,,
\end{equation}
where $l_I\,,b$ are constants that will turn out to be related to asymptotic moduli and
$p^0\,,q_I$ are the Taub-NUT charge and the electric charges of the solution, as computed
in the four dimensional theory. Then \eqref{feq}-\eqref{omeg_nab} are solved by:
\begin{eqnarray}
  f^{-1}X_I=L_I\,,\quad  \Rightarrow \quad f^{-3}&=&\frac 92 C^{IJK}L_IL_JL_K\,,\quad X^I=\frac 92\, f^2\, C^{IJK}L_JL_K\\
  \omega_5&=&-\frac bH\,,\qquad \hat\omega=0\,.
\end{eqnarray} 
As mentioned in section \ref{sec:5d}, we consider the case of a symmetric scalar manifold
which allows us to solve the scalar equation explicitly. These expressions agree with the
ones in \cite{0706.3373} and by using \eqref{bpsmetric2},\eqref{2form}, the full non-BPS
solution is reproduced, with the extra requirement of no Dirac-Misner strings. Note that
the extra rotation $b$ is invisible in the near horizon limit, which is described by the
supersymmetric attractor in \eqref{bps_flat}, as expected. In the following we concentrate
on the four dimensional interpretation of this solution.

When reduced to four dimensions using \eqref{red_ans}, the metric is the static limit of
\eqref{4dmetric}, with
\begin{eqnarray}\label{u_sing}
  e^{-4U}=\frac92 H C^{IJK}L_I L_J L_K - b^2\,, && \lim_{r\rightarrow\infty}e^{-4U}= h^0 l^3 - b^2=1,\\
  \quad l^3=l^{I}l_I,\quad &&l^I=\frac92 C^{IJK}l_Jl_K\,,\label{l_def}
\end{eqnarray}
where we imposed four dimensional asymptotic flatness and introduced some useful
notation. The four dimensional scalars \eqref{4d_scal} and their asymptotic values are
given by:
\begin{equation}\label{moduli_one}
  z^I = \frac{f\,X^I}{H}\left(b +\mathrm{i}e^{-2U} \right)\quad\Rightarrow\quad z^I_{\infty}\equiv x^I+\mathrm{i} y^I =\frac{l^I}{h^0l^3}(b+i)\,.
\end{equation} 
Thus, the parameter $b$ is identified as an asymptotic $B$-field in the four dimensional
theory.

The ADM mass associated to this four dimensional solution is simply found by expanding the 
metric  to first order in $1/r$. The result is:
\begin{eqnarray}
  M_{ADM} = \frac1{4G_4}\left(p^0 l^3 + 2\,h^0\, l^{I}q_I\right) 
  =\frac{\sqrt{1+b^2}}{4G_4} \left((1+b^2) y^{3/2} p^0 + \frac{2\,y^{I} q_I}{y^{3/2}}  \right)\,.
\end{eqnarray}
Here we replaced the $l_I$ by the four dimensional asymptotic scalars $y^I$, using the
constraint in \eqref{u_sing}, as well as:
\begin{equation}
  y^{3/2}=\sqrt{\frac16 C_{IJK} y^Iy^Jy^k},\qquad l^{3}=\sqrt{\frac16 C_{IJK} l^Il^Jl^k}\,,
\end{equation} 
where the second relation is an identity following from \eqref{eqn:symm}. This mass
formula is simply a sum of charges just like its BPS partner as anticipated above. It agrees
with the expression derived in \cite{0807.3503} for the special case of the $STU$ model.

In fact, the seed solution of \cite{0710.4967}, which was argued to have the minimal set
of parameters required to describe any non-BPS black hole in the STU model up to
$U$-duality is dual to the above solution for the STU case. More generally, the solutions
above have half of the charges turned on and asymptotic moduli as in \eqref{moduli_one}
for any scalar manifold. Based on this, we expect that the most general non-BPS black hole
for a four dimensional $\mathcal{N}=2$ theory with a symmetric scalar manifold can be
obtained by applying dualities on the above solution, along the lines of
\cite{0807.3503}. For more general scalar manifolds, one can still apply a restricted set
of dualities to generate solutions, like the electric-magnetic duality in
\cite{0706.3373}, but a case by case analysis is required.

A further property of this solution is that it is described by the first order formalism
of \cite{0702088,0706.0712,0706.3373}, as shown explicitly in \cite{0706.3373}. In
particular, the so called fake superpotential describing it can be obtained from the BPS
central charge using our recipe. In five dimensions, this amounts to a sign change for the
electric charges, while the four dimensional flow is governed by a function very similar
to the central charge. It follows that all non-BPS solutions that can be obtained by
$U$-duality from the above solution, must also exhibit this property. Again, for symmetric
scalar manifolds this can be done explicitly, following \cite{0807.3503}.

Finally, we discuss two possible generalisations. The first is to the multi-centre case,
by adding more centres in the harmonic functions $H\,,L_I$ in \eqref{har_fun_one},
generalising the base space to multi-Taub-NUT. Upon reduction to four dimensions, these
solutions seem to agree qualitatively with the solutions of \cite{0710.1638}, though we
have not considered them in detail. A second generalisation would be to turn on the
magnetic harmonic functions. It is important to notice that in the BPS case this would be
irrelevant due to the fact that only ratios of the form $K^I/H$ appear in the BPS
conditions, as observed in \cite{0504142}. In contrast, our equations
\eqref{attr}-\eqref{omeg_nab} do not appear to have such an invariance. In this case, the
functions $f\,, \omega_5$ diverge at the
centre\footnote{One would naively expect to find a horizon at $r=0$ in these coordinates,
  as in all other cases treated in this work.}
as $r^{-3}$ and $r^{-4}$ respectively, making them unattractive at first sight. However,
the near horizon geometry of these solutions is described by the BPS solutions considered
in \cite{0209114} (for the minimal theory). There, a number of curvature invariants were
examined and were found to remain finite at $r=0$, hinting at a regular solution. It would
be interesting to study this in more detail.

\subsection{A non-BPS two centre solution}
We now turn to the construction of a two centre example described by the almost BPS
partner of a supersymmetric black ring in Taub-NUT. Inspired by \cite{0504125}, we
dimensionally reduce this solution to obtain a two centre system in four dimensions with a
non-BPS black hole carrying $D4\!-\!D2\!-\!D0$ charges at one centre and a naked
singularity carrying $D6$ charge at the other. As in the BPS case, one can hide the naked
singularity by adding the electric black hole of the previous section at the centre of the
geometry, so that the solution becomes regular. It is worth mentioning that our solutions
seem to be related to the ones recently constructed in \cite{0811.2086,0811.2088}. In the
latter it was
argued that stationary non-BPS solutions similar to the ones in \cite{0504125} should
exist. It would be interesting to study this connection further.

Starting with the simple ring solution, we consider the set of harmonic functions:
\begin{eqnarray}\label{har_fun_two}
  K^I = \frac{2}{h^0} \frac{p^I}{|{\bf r- r}_0|},\quad
  L_I = l_I  +\frac2{3} \frac{q_{I}}{|{\bf r- r}_0|},\nn\\
  M = -b +  \frac{3}2  \frac{l_I p^I}{|{\bf r- r}_0|},\quad
  H= h^0+\frac{p^0}r,
\end{eqnarray}
where ${\bf{r}}_0=(0,0,R)$. The pole in the function $M$ is constrained by the requirement
of cancelling Dirac-Misner strings near asymptotic infinity through \eqref{omeg}. We
omitted the constant terms in the $K^I$ for simplicity, as they do not add important
features to the solution.

Unfortunately, the conditions \eqref{attr} and \eqref{omeg} cannot be solved explicitly in
the two centre case. Thus, the solution is only known implicitly through the first order
equations \eqref{attr}, \eqref{omeg} that govern the metric and the scalars and is unique
given the boundary conditions at infinity implied by \eqref{har_fun_two}. This brings two
complications in our analysis. First, we don't have complete control of global issues like
closed time-like curves (CTC's) or Dirac strings in the full geometry. We will assume that
there are no such obstructions for this solution, as for its BPS partner. In the following
we verify that there are no such problems at least asymptotically and near the horizon, by
restricting the charges and moduli accordingly. It appears that by requiring the
conditions
\begin{equation}
  e^{-4U}\geq 1\,,\qquad \hat\omega^i\hat\omega_i < e^{-4U}\,,
\end{equation}
where the function $e^{-4U}$ was defined in \eqref{4dmetric}, one can eliminate these
problems for the four dimensional solution that interests us most. As $e^{-4U}$ is
proportional to $g_{\psi\psi}$ in the full geometry, these constraints are also natural
from a five dimensional point of view.

Another implication of the difficulty in solving for the full flow is that we are not able
to find directly the relation of the distance $R$ to the charges and moduli, since this
would be obtained by an integrability condition on \eqref{omeg} after \eqref{attr} and
\eqref{omeg_nab} are solved. Thus we will be forced to work indirectly, comparing with
known attractor formulae in four dimensions to determine this parameter.

Here, we will again concentrate on the dimensionally reduced solution, for which
\eqref{attr}-\eqref{omeg_nab} describe a first order split attractor flow for all
quantities and especially the scalars, as in the BPS case. Explicitly, the metric of the
four dimensional solution is as in \eqref{4dmetric}, whereas the scalars (and their
asymptotic values) and field strengths are (assume that the scalars take values in a
symmetric space for simplicity):
\begin{eqnarray}\label{scal_two}
  z^I
  =-\frac{f}{H}\,X^I\,\left(H\,\omega_5- \mathrm{i}e^{-2U}\right) 
  + \frac12 K^I \quad\Rightarrow\quad z^I_{\infty}\equiv x^I
  + \mathrm{i}\,y^I = \frac{b\,l^I}{h^0\,l^3}\left(1+\mathrm{i}\right)\,,
\end{eqnarray}
\begin{eqnarray}
  F^I_{(4)}&=&-d\left[\left( X^If(1+H^2\omega_5^2\,e^{4U})+\frac12 K^IH^2\omega_5e^{4U}\right)(dt+\hat\omega)\right] \nn\\
  &&+\frac14 \left(K^I\nabla_kH - H\nabla_kK^I\right)\varepsilon_{ijk}dx^i\wedge dx^j\,,
\end{eqnarray}
and $F^0=dA^0$, with $A^0$ as in \eqref{4dKK}. Note that the asymptotic scalars are as in
the single centre case, but they can be generalised by adding constant terms in the
harmonic functions $K^I$.

We now discuss the behaviour of the solution near infinity and the horizon. This will
enable us to calculate the mass from asymptotic data and the entropy through the horizon
area of the black ring. Consider first the limit $r\rightarrow \infty$, so that all the
harmonic functions can be approximated by powers of $1/r$. Then, the solutions of
\eqref{attr},\eqref{omeg_nab} up to first order in that expansion are:
\begin{eqnarray}
  f^{-1}X_I\simeq L_I\quad\Rightarrow \quad H \omega_5 \simeq \frac32 \frac{l_Ip^I}{r}+ M = -b +  \frac {3\,l_Ip^I} r ,
\end{eqnarray}
\begin{equation}\label{as_flat}
  \lim_{r\rightarrow \infty}e^{-2U}=\sqrt{h^0l^3 - b^2  }=1
\end{equation}
and the asymptotic solution reduces to the one in the single centre case with the addition
of magnetic charges. Note that we imposed the same constraint on the moduli. We will use
the definitions in \eqref{l_def} in the following. The mass of the four dimensional two
centre solution is found again by expanding $e^{-2U}$, with the result:
\begin{eqnarray}
  M_{ADM} &=& \frac{1}{4G_4}\left(l^3 p^0+ 2\,h^0\, l^I q_{I} + 6 b l_Ip^I \right) \nn\\
  &=&\frac{\sqrt{1+ b^2}}{4G_4} \left((1+b^2) y^{3/2} p^0 + \frac{2\,y^{I} q_I}{y^{3/2}} + 
    \frac{6 b}{y^{3/2}} y_Ip^I \right)\,,\label{4d_mass}
\end{eqnarray}
where we wrote everything in terms of the asymptotic moduli and we defined
$y_I=\frac16 C_{IJK}\,y^J\,y^K$. We observe again that it is a marginal sum of charges, as
expected, implying that there is no binding energy between the constituents of the
system. We will have more to say on this when we consider the distance between the two
centres.

Following \cite{0504125}, we also compute the ADM momentum along the fifth direction,
identified as electric Kaluza-Klein charge after dimensional reduction:
\begin{eqnarray}
  P_{ADM}
  &=& -\frac1{16\pi\,G_5}\int d\bar \psi\, d \Omega_{2}\,r^2\,\partial_r\,h_{\bar t\bar \psi}\nn\\
  &=& \frac{\pi}{G_5}
  \left[
    -\frac{2\,\tilde b}{l}\,l^Iq_I + \frac{\tilde b^2}{h^0}p^0 - \frac{3}{\sqrt{1+
        b^2}}\,l_Ip^I \left( 1+\tilde b^2\left(1+\frac12 l^2\right) \right)
  \right]\,,
\end{eqnarray}
where $h_{\mu\nu}=g_{\mu\nu}-\eta_{\mu\nu}$ and the $\bar t\,,\bar \psi$ coordinates are
related to $t\,,\psi$ by a Lorentz boost, so that
$\lim_{r\rightarrow \infty}g_{\,\bar t\bar \psi}=\eta_{\,\bar t\bar \psi}$, as in
\cite{0504125}.  We also set $ \tilde{ b} \equiv b/\sqrt{h^0l^3} =b/\sqrt{1+b^2}$, in view
of \eqref{as_flat}. In four dimensions, the charge associated to the $A^0$ gauge field
differs from this result, as it is defined through the electric-magnetic dual of the
corresponding field strength, as in \eqref{el_chsr}. The charges of our four dimensional
solution, as computed from \eqref{el_chsr}-\eqref{mag_char} are:
\begin{eqnarray}\label{4d_char}
  q_A=\left(-\frac{3\,l_Ip^I}{2h^0}\,, - q_I\right)\,,\qquad p^A=(p^0\,,p^I)\,,
\end{eqnarray}
where the $q_I\,,p^I$ are the ones appearing in \eqref{har_fun_two}. Note that the charge
$q_0$ is proportional to the pole of the function $M$, as would be naively expected.

As a final remark on the asymptotic properties of this solution, it is important to
emphasise that the electric charges in five dimensions are not equal to the four
dimensional ones, as has been observed in the BPS case \cite{0504142,0408122,0506251}. As
can be easily checked, the five dimensional electric charges are shifted with respect to
the $q_I$ by a factor proportional to $b\,C_{IJK}l^Jp^K$ coming from the Chern-Simons
term. It would be interesting to analyse the implications of this inconspicuous moduli
dependent shift on both the macroscopic supergravity solutions and the microscopic
counting of the entropy for the black ring.

Now, we turn to the near horizon region of the ring, taking the limit
${\bf r}\rightarrow{\bf r_0}$. Fortunately, the near horizon region is identical to the
one of the BPS solution, as explained above, so that most results of
\cite{0504125,0408122} carry over by simply replacing $p^I\rightarrow H_0\,p^I$, where
$H_0=H({\bf r_0})$. In particular, the metric of a cross section of the horizon is:
\begin{equation}
  ds^2_\mathrm{hor} =  L^2 d{\psi}^2 + \left(\frac{H_0}{h^0}\right)^2\,p^2 \left(d\theta^2 + \sin^2 \theta d\phi^2\right)\,,
\end{equation}
where $0\leq\theta\leq\pi$, $0\leq\phi\leq 2\pi$ are spherical coordinates on the ring and
$0\leq\psi\leq 4\pi$ is parametrising the $S^1$ of Taub-NUT along which the ring is
wound. The constants $p$, $L$ are given by:
\begin{eqnarray}
  p^3=\frac 16 C_{IJK}p^Ip^Jp^K\,,\qquad L^2=\frac{2\,(h^0)^2}{H_0^2\,p}\left(\,D^{IJ}\,q_I\,q_J + q_0 \right)\,, \qquad D^{IJ}\,C_{JKL}p^L= \delta^I_K\,,
\end{eqnarray}
with $q_0$ as in \eqref{4d_char}. Requiring that $L^2>0$ eliminates CTC's near the
horizon, as in \cite{0504125}.

The entropy is found through the area of the horizon:
\begin{eqnarray}\label{ring_entr}
  S=\frac A{4\,G_5}
  =\frac\pi{G_4}\,L\,\left(\frac{H_0}{h^0}\right)^2\,p^2= \frac{H_0}{h^0}\, \frac{2\pi}{G_4} \sqrt{\frac{p^3}2\left(\,D^{IJ}\,q_I\,q_J + q_0 \right)}\,,
\end{eqnarray}
where we use four dimensional quantities. This expression agrees up to the prefactor with
standard entropy formulae for the prepotential \eqref{prepotential} for both BPS and
non-BPS attractors obtained by analytic continuation of charges in four dimensions. As can
be easily seen from \eqref{scal_two}, the same holds for the scalars and gauge fields with
exactly the same prefactor. We use this to fix the parameter $R$,
comparing\footnote{Note that we use the conventions of \cite{0007195}} with the
expressions in \cite{0511117}. We find that:
\begin{eqnarray}
  \frac{H_0}{h^0}\equiv 1+ \frac{p^0}{h^0R}=1\,,\quad\Rightarrow\quad R\rightarrow\infty\,,
\end{eqnarray}
which appears puzzling at first sight. Recalling that one finds the same result from
\eqref{omeg_BPS} for the distance between two centres carrying mutually local charges in
the BPS case, it suggests that such a divergence simply signals that the distance is not
fixed, but arbitrary. In other words, the near horizon limit imposes $R\rightarrow\infty$,
as this quantity is not fixed in terms of the charges that have a definite scaling as one
takes the limit. We view this as more evidence towards the marginality of the system,
consistent with the mass formula derived above. It remains an open question to solve
\eqref{attr}-\eqref{omeg_nab} (numerically or otherwise) and verify this result in the
full solution.

A related observation is that the non-BPS attractor carrying all possible charges in four
dimensions could be unstable, hinting towards the existence of a multi-centred solution
with the same total charge and stable components \cite{0606263,0511117,0705.4554}. Our
solution (more precisely the one described below, where both centres are proper
attractors) is indeed a resolution of such a system.  Moreover, the marginality of its
constituents seems to indicate that the single centre attractor would be stable as well,
unless this limit lies outside the boundary of moduli space of these solutions. This
brings us back to the discussion in the end of section \ref{sing_exa}, since all harmonic
functions would be sourced at one centre in this case. It remains an interesting open
question to study these single centre geometries both in five and four dimensions and
investigate whether they are proper attractors.

This completes our discussion of the non-BPS black ring and we now briefly sketch how to
obtain a regular solution in four dimensions. As mentioned above, reducing a Taub-NUT
solution to four dimensions produces a naked singularity carrying Kaluza-Klein magnetic
charge $p^0$, even though the full five dimensional solution is smooth. One can construct
a regular solution by hiding the naked singularity behind the horizon of a black hole
placed at the centre of the five dimensional geometry, as in \cite{0504125}. This is
easily done through the following modification on the electric harmonic functions:
\begin{eqnarray}
  L_I = l_I +\frac2{3} \frac{q_{I}^{(h)}}{r} +\frac 2{3} \frac{q_{I}}{|{\bf r- r}_0|}\,,\qquad
  M = -b +  \left(\frac{3}2 l_A p^A - q_0^{(r)} \right) \frac 1{r}+ \frac {q_0^{(r)}}{|{\bf r- r}_0|}\,.
\end{eqnarray}
This turns the four dimensional naked singularity into a rotating version of the black
hole in the previous section, so that both centres are regular black holes with finite
area horizons. Note that now there is an extra $D0$ charge present, unlike in previous
examples. It can be assigned to any of the two centres, with the constraint that the total
$D0$ charge is as in the simple ring solution, due to the requirement of cancelling
Dirac-Misner strings asymptotically.

All the above results remain valid after this modification, by simply replacing $q_0$ by
$q_0^{(r)}$ in the entropy formula \eqref{ring_entr} for the ring and $q_I$ by
$q_I+ q_{I}^{(h)}$ in all other expressions. The total electric charge $q_0$ is the same
as in \eqref{4d_char}. In particular, the mass formula \eqref{4d_mass} remains marginal,
implying that there is no interaction energy between the two centres even after adding
extra charges.

We end this section with a few comments on possible generalisations. Comparing with the
most general two centre solution in four dimensions, we see that our regular solution has
half of the charges turned on at one centre, all but one charges at the other and
asymptotic moduli as in \eqref{scal_two}. Restricting to symmetric scalar manifolds, we
see that this can be dualised to arbitrary moduli and charges at one centre, as in the
single centre solution above. We are then only one charge short of the most general two
centre solution. Finally, it should be straightforward to construct multi-black rings along 
the lines of \cite{0408122}. These solutions would reduce to a collection of magnetically 
charged black holes in four dimensions.

\section{Conclusion and Discussion}
\label{sec:conc}

In this work, we have identified a systematic map from BPS to non-BPS extremal black holes
in four dimensional $\mathcal{N}=2$ supergravity with cubic prepotential, that generalises
the one known through the study of attractors. In doing this, we found that a simple
algorithm can be obtained in the five dimensional uplift of the theory, by exploiting the
properties of an extra set of BPS solutions available for a flat base space. These
solutions are invariant under a different $SU(2)$ isometry than the one seen in four
dimensions and are invisible in the four dimensional BPS spectrum. By gluing together such
local BPS solutions with flat base, we constructed solutions with general Gibbons-Hawking
base space that preserve half of the supersymmetry locally but not globally, nicknamed almost BPS. After dimensional
reduction, the result is a class of non-BPS solutions very similar to the supersymmetric
solutions that have been studied extensively. In particular, all quantities in these
solutions satisfy first order equations, so that a number of important techniques
developed for the supersymmetric case can be applied. We have demonstrated this by
constructing multi centred solutions through arbitrary harmonic functions, following
similar work on BPS solutions \cite{0408122,0209114}.

Our class of almost BPS solutions is as large as the BPS class, in the sense that it
involves the same number of harmonic functions, as described below \eqref{omeg_nab}. It
includes the non-BPS solutions obtained through analytic continuation of charges
\cite{0511117, 0606263, 0705.4554} and some recent generalisations
\cite{0706.3373, 0710.4967}. Thus, it is bigger than the set of non-BPS solutions known to
date (up to U-duality), but smaller than the full class of non-BPS solutions. As explained
in section \ref{noBPS}, it only includes solutions very closely related to BPS solutions,
representing special points in the four dimensional non-BPS duality orbits. However, in
cases with enough symmetry, one can use our special solutions as seeds to generate the
full orbit of non-BPS solutions, as in \cite{0807.3503}.

The defining property of our solutions, that they preserve four supercharges locally, was
also used to clarify previous observations on special properties of four dimensional
attractors \cite{0701176, 0707.0964} and solutions \cite{0710.4967}. As both the
asymptotic region and the near horizon geometry of extremal black holes can be written as
a timelike fibration over flat space, our five dimensional solutions are BPS in these
regions. This crucial property implies a simple marginal mass formula for all such
solutions, as in section \ref{sec:eg}, and simplifies near horizon considerations. In this
special setting, we have analysed a known single centre example and a new two centre
solution in four dimensions that represents a resolution of a possibly unstable non-BPS
attractor \cite{0606263, 0705.4554}. In both cases the mass formula suggests a marginally
bound state, and this conclusion is strengthened by the fact that the distance between the
two centres in the second case appears to be arbitrary at the level of our approximation.

There are several directions to be investigated in the context of our solutions. First, it
seems very likely that our almost BPS conditions \eqref{attr}-\eqref{omeg_nab} can be
rewritten in terms of the symplectic section $(Y^A,\,\,F_A)$, that is the more natural
notation in four dimensions. This could shed more light on the systematics of these
solutions from a four dimensional perspective and perhaps help towards a plausible
generalisation of this construction to arbitrary prepotentials. Moreover, such a
formulation would make the properties of our solutions under four dimensional electric
magnetic duality manifest \cite{0007195, 0103086, 0807.4039}, simplifying the construction
of new solutions through dualities. Such a purely four dimensional investigation is also
required to identify exactly what subspace of non-BPS solutions is spanned by the almost
BPS solutions and their duals.

More importantly, it would be very interesting to investigate whether the marginality
property of our examples is generic or not. Thus, one could distinguish the cases where a
nontrivial moduli space of this restricted class of solutions exists, if any. The general
lore is that the intricate moduli spaces of the BPS solutions should not be present, but
there could still be interesting questions to be asked for a trivial moduli space, for
example what happens at its boundary. A related, but equally challenging task would be to
study the properties of the split attractor flows described by the almost BPS
conditions. Such a program appears to require heavy use of numerical methods, in view of
the extra complications compared to the BPS case. In fact, it should be noted that such
studies could be problematic in view of the incompleteness of our set of solutions
relative to the full non-BPS spectrum.

Finally, it would be interesting to check the robustness of our trick against higher
derivative corrections. The most crucial property of the two derivative theory we used
is the hyper-K\"ahler structure of the base. Recently, the $R^2$
corrected five dimensional theory was considered in \cite{0703087, 0705.1847, 0801.1863},
where some explicit solutions were constructed. These were found to be timelike fibrations
over hyper-K\"ahler spaces, so that a possible extension of our analytic continuation to
this theory is conceivable. Such an extension could allow the construction of non-BPS solutions
in four dimensional $\mathcal{N}=2$ supergravity with higher derivatives
\cite{0009234, 0012232, 0007195, 0603149}. We hope to return to some of these issues in
the future.

\acknowledgments{It is a pleasure to acknowledge fruitful discussions with
G. L. Cardoso and B. de Wit in various stages of this work and on an earlier version of
this manuscript. We would also like to thank J. M. Oberreuter and S. Vandoren for useful
discussions. The work of S.K. is part of the research program of the `Stichting voor
Fundamenteel Onderzoek der Materie (FOM)', which is financially supported by the
`Nederlandse Organisatie voor Wetenschappelijk Onderzoek (NWO)'. This work has been partly
supported by EU contracts MRTN-CT-2004-005104 and MRTN-CT-2004-512194, INTAS contract
03-51-6346, and by NWO grant 047017015.}

\begin{appendix}
\section{The equations of motion} \label{feom}

In this appendix, we outline how both expressions in \eqref{bpsmetric2}-\eqref{lapl} solve
the equations of motion. The equations of motion found by varying the action
\eqref{5dact} with respect to the metric, gauge fields and the $n_v-1$ independent scalars
$\chi^a$ obtained by solving the constraint in \eqref{conda}, respectively can be written as:               :
\begin{eqnarray}
\label{eqn:mot}
R_{\mu \nu} &+&Q_{IJ} F^I{}_{\mu \lambda} F^J{}_\nu{}^\lambda
+Q_{IJ} \partial_\mu X^I \partial_\nu X^J-\frac16 g_{\mu \nu}
Q_{IJ} F^I{}_{\kappa\lambda} F^{J \kappa\lambda}  =0\,,\nonumber\\
&&d \left(Q_{IJ} \star F^J \right)+\frac14 C_{IJK} F^J \wedge F^K=0\,,\\
\bigg[d ( \star dX_I )&+& \bigg( X_J X^L C_{IKL}-\frac16
C_{IJK} \bigg) (F^J \wedge  \star\, F^K +dX^J \wedge \star\, dX^K )\bigg]
\frac{\partial X^I}{\partial \chi^a} = 0\,,\nonumber
\end{eqnarray}
where $\star$ is the Hodge dual in five dimensions. Here and in the following, we use the identities:
\begin{equation}
Q_{IJ} X^J = \frac32 X_I\,,
\qquad
Q_{IJ}\, \frac{\partial X^J}{\partial \chi^a} = -\frac32 \frac{\partial X_I}{\partial \chi^a}\,,
\end{equation}
which are implied by \eqref{conda}-\eqref{scamatr}.

We consider a metric ansatz as in \eqref{bpsmetric2} and we take the gauge fields to be of
the form
\begin{eqnarray}
F^I=\pm \,d(fX^I)\wedge (dt+\omega) \pm X^IG + \Lambda^I
\end{eqnarray} 
where, for the moment, the $\Lambda^I$ are arbitrary closed two-forms on the
hyper-K\"ahler base and as before, $G=fd\omega$.  It is also convenient to define 
\begin{equation}
  \label{ji_def}
  J^I = \pm X^IG + \Lambda^I.
\end{equation} 
The trace-reversed Einstein equations then take the form:
\begin{eqnarray}\label{einst}
&&X^I\Delta( f^{-1}X_I)-\tfrac14 ( G^{2}- \tfrac23 Q_{IJ}J^{I}\cdot J^J)=0\,,\nonumber\\
&&f\,\nabla^qG_{pq} + \nabla^qf \left(2G_{pq}-3\,X_IJ^I_{pq}\right) +
3\,\nabla^q X_I\, J^I_{pq} =0\\
&& \tfrac12 \,\left[X^I\Delta( f^{-1}X_I)-\tfrac14 ( G^{2}- \tfrac23 Q_{IJ}J^{I}\cdot J^J) \right]\delta_{p q}  - G^+_{pr} G^-_q{}^r + Q_{IJ} \, J^{I+}_{pc}J^{J-}_q{}^c\nonumber=0\,,
\end{eqnarray}
where the identity 
\begin{equation}
A^{\pm pr}\,B^\pm_r{}^q + A^{\pm qr}\,B^\pm_r{}^p\, =\, \ft12 \delta^{pq}\, 
A^{\pm rt}\,B^\pm_{rt}\,\equiv\, \ft12 \delta^{pq}\, A^{\pm}\cdot B^\pm
\end{equation}
was used. Here, Latin letters are used for flat indices on the base space and the superscript $\pm$ denotes 
(anti-)selfdual forms on it.
In order to solve these equations, we choose
\begin{equation}
X_I\Lambda^I=\mp\frac23 G^{\pm} \,,
\end{equation}
as in \eqref{bpsl2}. Taking a derivative and using that the $\Lambda^I$ are closed leads to: 
\begin{equation}\label{eom_id}
 \nabla^q G_{pq}= \mp f^{-1} (\tilde{*}G)_{pq} \nabla^q f - 3\,(\tilde{*}\Lambda^I)_{pq} \nabla^q X_I\,,
\end{equation}
where $\tilde{*}$ denotes the Hodge dual on the base. It is then easy to show that the second and 
third equations in \eqref{einst} are satisfied if the first one is. 

Similarly, the Maxwell equation in \eqref{eqn:mot} implies the two equations:
\begin{eqnarray}
\pm \tfrac32\, d\tilde{*}d( f^{-1} X_I)+ 3\,X_I G\wedge \left( G^\pm+ \tfrac32 X_I\tilde{*}\Lambda^I \right)&&\nonumber \\
\pm C_{IJK}X^K \,G\wedge \Lambda^{I\mp} +\tfrac14 C_{IJK}\Lambda^J\wedge\Lambda^K &=&0\,, 
\end{eqnarray}
\begin{eqnarray}
2\,f^{-1}df\wedge \left(\tfrac12 X_I G\pm \tfrac13 C_{IJK} X^K \Lambda^{\mp\,J} \right)
+ 3\, X_IdX_J\wedge\tilde{*}\Lambda^J&&\nonumber\\
+2\,f^{-1}d(fX_I)\wedge(G^\pm+ \tfrac32 X_I\tilde{*}\Lambda^I) \pm X_I d\tilde{*}G&=&0\,.
\end{eqnarray}
When the above choices for the $\Lambda^I$ are imposed, a bit of algebra shows that the first reduces to:
\begin{equation}\label{lapl2}
d\,\tilde{*}\,d( f^{-1} X_I)=\mp \tfrac16 C_{IJK}\Lambda^J\wedge\Lambda^K\,,
\end{equation}
which can be shown to imply the $00$ component of the Einstein equation above, while the second is identically satisfied by using \eqref{eom_id}.

Finally, the scalar equation in \eqref{eqn:mot} reduces to:
\begin{eqnarray}
\label{eqn:scal}
\bigg[\Delta(f^{-1} X_I)  \pm \frac13 X^P C_{NPI}G\cdot \Lambda^N\qquad&&\\
+\frac12\bigg( X_J X^L C_{IKL}&-&\frac16 C_{IJK} \bigg)\Lambda^J\cdot\Lambda^K\bigg]
\frac{\partial X^I}{\partial \chi^a} = 0\,.\nonumber
\end{eqnarray}
For the above $\Lambda^I$, the expression in brackets leads again to \eqref{lapl2}. 
\end{appendix}

\bibliography{4d5d} \bibliographystyle{JHEP}
\end{document}